\documentclass{article}
\usepackage{spconf,amsmath,graphicx}
\usepackage{comment}
\usepackage{subfigure}
\usepackage{booktabs}
\usepackage[dvipsnames]{xcolor}

\def\etal{{et al. }}
\title{DISENTANGLEMENT FOR AUDIO-VISUAL EMOTION RECOGNITION USING MULTITASK SETUP}

\name{Raghuveer Peri$^{\dagger}$ \sthanks{Work was completed while interning at Amazon} \qquad Srinivas Parthasarathy $^{\star}$ \qquad Charles Bradshaw$^{\star}$ \qquad Shiva Sundaram$^{\star}$}

\address{$^{\dagger}$ University of Southern California, Los Angeles, CA, USA \\
	$^{\star}$Amazon Inc., Sunnyvale, CA, USA}
\begin{document}
\ninept
\maketitle
\begin{abstract}
Deep learning models trained on audio-visual data have been successfully used to achieve state-of-the-art performance for emotion recognition. In particular, models trained with multitask learning have shown additional performance improvements. However, such multitask models entangle information between the tasks, encoding the mutual dependencies present in label distributions in the real world data used for training. This work explores the disentanglement of multimodal signal representations for the primary task of emotion recognition and a secondary person identification task. In particular, we developed a multitask framework to extract low-dimensional embeddings that aim to capture  emotion specific information, while containing minimal information related to person identity. We evaluate three different techniques for disentanglement and report results of up to 13\% disentanglement while maintaining emotion recognition performance.
\end{abstract}
\section{Introduction}
\label{sec:intro}

Emotions play an important role in human communication. Humans externalize their reactions to surrounding stimuli through a change in the tone of their voice, facial expressions, hand and body gestures. Therefore, automatic emotion recognition is of interest for building natural interfaces and effective human-machine interaction. \cite{pantic2003toward}. With regards to human communication, emotion is primarily manifested through speech and facial expressions, each providing complementary information \cite{mehrabian2008communication}. Therefore, multimodal techniques have been widely used for reliable emotion prediction  \cite{kim2013deep,wang2008recognizing,song2004audio}. 

Several studies have shown that emotion recognition benefits from training with secondary related tasks through multitask learning (MTL). In Parthasarathy and Busso \cite{parthasarathy2017jointly}, predicting the continuous affective attributes of valence, arousal and dominance are treated as the multiple tasks and trained jointly. In Li \etal \cite{li2019improved} and Kim \etal \cite{kim2017towards}, gender prediction as a secondary task improves emotion recognition performance by upto 7.7\% as measured by weighted accuracy on a standard corpus. A more comprehensive study involving domain, gender and corpus differences was performed in Zhang \etal \cite{zhang2017cross}, where cross-corpus evaluations showed that, in general, information sharing across tasks yields improvements in performance of emotion recognition across corpora. These studies indicate that several paralinguistic tasks help generalize shared representations that improve overall performance of the primary task. This motivates us to use person identification as a secondary task to help improve performance on the primary emotion task.

With MTL the shared representations among tasks retain information pertaining to all the tasks. While this generalizes the overall architecture, it does so by entangling information between multiple tasks  \cite{liang2020model, xiao2018learning, williams2019disentangling}. 
Since most machine learning models are trained on human-annotated, unconstrained real-world data, several factors that should theoretically be independent end up being dependent. For e.g. in the case of emotions, studies have shown the correlation with demographical information \cite{chaplin2015gender}. Therefore, MTL inherently captures the joint dependencies between different factors in the data. This is problematic as the gains through generalization across tasks may lead to bias and subsequently poor performance on unseen data.


To address the entanglement of information in MTL, this paper develops a multimodal emotion recognition model, improves its performance using person identification as a secondary task and subsequently disentangles the learned person identity information, while still maintaining the improved emotion recognition performance. As an additional contribution, we analyze how much emotion information is present in the identity representations when models are trained in a MTL setup. For disentanglement, we experiment with three distinct disentanglement techniques to minimize the information transfer between speaker embeddings and emotional labels and vice-versa. We present experiments that make use of alternate adversarial training strategy, gradient reversal based technique adapted from Domain Adversarial Training~(DAT) literature and a confusion loss based technique inspired from \cite{alvi2018turning}.
We evaluate the models pre and post disentaglement, showing that disentanglement retains or improves performance on primary tasks upto 2\% absolute, while reducing the leakage of information between the tasks with disentanglement upto 13\% as measured by F-score.


\section{RELATED WORK}
\label{sec:related}

In the context of representation learning for emotion recognition, the goal is to extract low dimensional embeddings that are invariant to factors such as domain and speaker. Abdelwahab and Busso \cite{abdelwahab2018domain} used gradient reversal (GR) to extract emotion representations that are invariant to domain. \textcolor{black}{Mao et al.~\cite{mao2017learning} imposed an explicit orthogonality criterion to encourage the learning of domain invariant and emotion discriminative features.}
Similarly, to extract speaker-invariant emotion representations, adversarial learning approach was explored in addition to an online data augmentation technique by Tu et al.~\cite{tu2019towards}. They showed improvements in the emotion recognition performance while testing on speakers unseen during training. More recently Li et al. \cite{li2020speaker}  proposed an entropy-based loss function along with GR and showed improved performance compared to \cite{tu2019towards}. Kang et al. \cite{kang2020disentangled} propose channel and emotion invariant speaker embeddings. However, most of these works consider emotion recognition using speech modality alone. \textcolor{black}{Jaiswal and Provost \cite{jaiswal2020privacy} explored privacy-preserving multimodal emotion representations, where audio and text modalities were utilized.} Our study differs from previous studies by using a secondary task to improve primary emotion recognition performance while being invariant to the auxiliary factors.

With regards to identity embeddings, Wiliams and King \cite{williams2019disentangling} have shown that speaker embeddings capture significant amount of affect information. It has been found that differences in the affective states of a person between training and testing conditions can degrade the performance on the task of identity verification from speech \cite{parthasarathy2017study, wu2006study}. Techniques have been proposed to compensate this by transforming features from expressive speech to neutral speech domain \cite{bao2007emotion,krothapalli2012neural}. While most of the existing works learn identity representations separately and then try to make them invariant to emotional states, we co-learn identity representations with an emotion recognition task while simultaneously removing emotion information from them.



\section{METHODOLOGY}
\label{sec:method}

\begin{figure}[htb]
	\centering
	\centerline{\includegraphics[width=8cm]{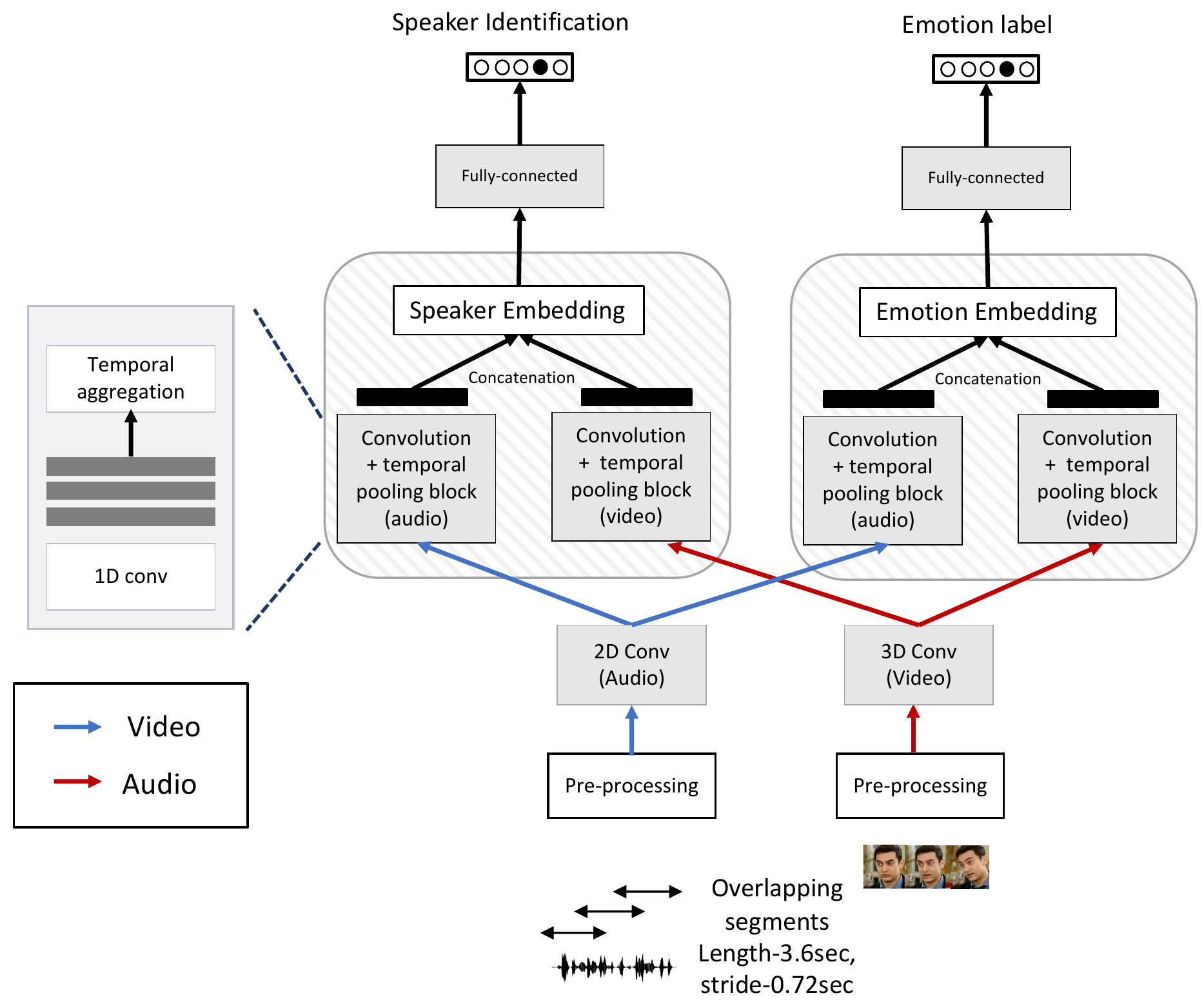}}
	
	%
	\caption{Block diagram depitcing the baseline multimodal, multitask training}
	\label{fig:MTL}
\end{figure}

\begin{figure}[htb]
	\centering
	\centerline{\includegraphics[width=8cm]{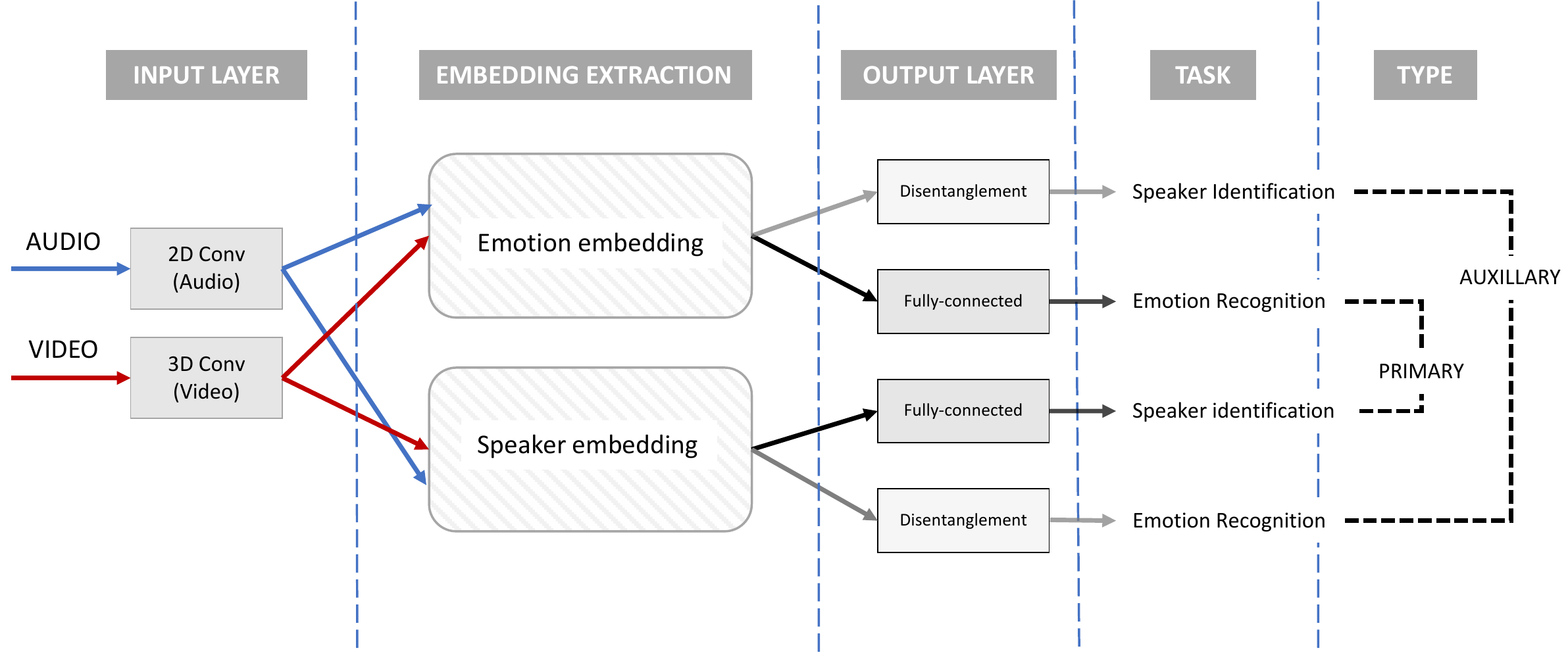}}
	
	%
	\caption{Block diagram depicting the baseline model with an auxiliary disentanglement task}
	\label{fig:auxiliary}
\end{figure}

Fig. \ref{fig:MTL} illustrates the multitask architecture for emotion recognition and person identification. The inputs to the model are both audio and video frames that are time-synchronized. The first step is a shared convolutional feature extraction stage where a data-driven representation is extracted for both audio and video independently. The architectures for this first stage blocks are adopted from \cite{nagrani2020disentangled}. A second level temporal aggregation block pools the feature representation for audio and video separately over entire clips to fixed dimensional representation. The outputs of the audio and video pooling blocks are concatenated; resulting in independent embedding layers \emph{emotion embedding} and \emph{speaker embedding}. The final output layers for task-specific outputs are fully connected layers with a softmax activation function to predict the emotion and person identity labels respectively.  Please note that we have used the terms \emph{speaker identity} and \emph{person identity} interchangeably throughout the paper.

Fig. \ref{fig:auxiliary} illustrates the addition of auxiliary branches to the baseline multitask architecture. The auxiliary branches are used to assess the amount of emotion information in the speaker embeddings and vice versa. These auxiliary branches are also used for disentanglement as explained in Section \ref{sssec:auxiliary}.

\subsection{Pre-processing}
\label{sssec:pre_proc}
{\color{black}
The input audio and face crop streams from a video clip are first fed into corresponding pre-processing blocks. On the audio stream, pre-processing includes extracting log Mel frequency spectrogram features on overlapping segments of fixed length and stride. This results in one feature vector per segment, with varying number of segments per video clip, depending on the length of the clip. In order to perform efficient batch processing, we pad the features with a constant value to ensure that each video clip contains the same number of segments, $N$. The resulting features have the dimensions $B*N*D_a$ where $B$ is the minibatch size and $D_a$ is the dimension of the Mel spectrogram features.
On the face crops, pre-procesing includes resizing them into a fixed size of $D_v*D_v$ pixels and rescaling the values to between $-1$ and $1$. The resulting face crops have the dimensions $B*N*D_v*D_v$.}


\subsection{Auxiliary branch for disentanglement}
\label{sssec:auxiliary}
The multitask outputs are built on top of the common embedding layers for the emotion and person identification tasks respectively.  As a result, when training the model, it tends to train an entangled embedding that is optimized for both tasks. This form of entanglement could lead to learning needless dependencies in the train set that may affect the overall generalization. In this work, both for person identification and emotion recognition, the second output or auxiliary task is used to disentangle the emotion information from the speaker embeddings and vice-versa (Fig. \ref{fig:auxiliary}).
Disentanglement is achieved using the auxiliary branch. The basic intuition here  is similar to domain adversarial training, where the goal is to learn representations that are optimized for the primary task, while simultaneously training it to perform poorly on the auxiliary  task. To this end, we experiment with three techniques for disentanglement: (1) gradient reversal, (2) alternate primary-auxiliary training and (2) and confusion loss~(CONF).

\textbf{Gradient reversal} was originally developed in Ganin and Lempitsky \cite{ganin2015unsupervised} to make digit recognition task invariant to domain through adversarial training. As discussed in Section \ref{sec:related}, it was adapted to extract speaker-invariant speech emotion representations in Tu \etal \cite{tu2019towards}. Gradient reversal is achieved  by introducing it in the stages of a network where the auxiliary branch separates from the primary branch. This layer has no effect in the forward pass of training, while in the backward pass the gradients from the auxiliary branch are multipled by a negative value before  backpropagating  it to the embedding layer.

\textbf{Alternate training} strategy for disentanglement was inspired from adversarial training literature \cite{goodfellow2014generative}, where two models are trained with competing objectives. In our setup, for emotion embeddings, the primary task is to predict the emotion labels, while the auxiliary task is to predict person identity labels. Equations \ref{eq:eq1} and \ref{eq:eq2} show the loss functions of the primary and auxiliary branch respectively, which are modeled as cross-entropy loss.
$\hat{{e}}_{prim}$ and $\hat{{s}}_{prim}$  denote the primary predictions from the emotion and speaker identification branches respectively. Similarly, $\hat{{e}}_{aux}$ and $\hat{{s}}_{aux}$ denote the auxiliary predictions from the speaker identification and emotion recognition branches respectively. $e_{target}$ and $s_{target}$ denote the groundtruth emotion and speaker identity labels.
\begin{equation}
\begin{split}
L_{primary} = & w_{em\_prim} * L(\hat{{e}}_{prim}, {e_{target}}) \\
& + w_{spk\_prim} * L(\hat{{s}}_{prim}, {s_{target}})
\label{eq:eq1}
\end{split}
\end{equation}
\begin{equation}
\begin{split}
L_{auxiliary} = & w_{spk\_aux} * L(\hat{{e}}_{aux}, {e_{target}}) \\
& + w_{em\_aux} * L(\hat{{s}}_{aux}, {s_{target}})
\label{eq:eq2}
\end{split}
\end{equation}

Alternate training proceeds in a minimax fashion. The auxiliary branch is trained to minimize $L_{auxiliary}$, while the primary branch is trained to minimize $L_{primary}$  and simultaneously maximize $L_{auxiliary}$.

\textbf{Confusion loss} for disentanglement has been introduced in Tzeng \etal \cite{tzeng2015simultaneous} and adapted for disentangling person identity and spoken content representations in Nagrani \etal \cite{nagrani2020disentangled}. We apply a similar strategy to disentangle the emotion and person identity representations. On a high level, the loss forces the embeddings such that, for the auxiliary task, each class is predicted with the same probability. Similar to \cite{nagrani2020disentangled}, we implement the confusion loss as the cross-entropy between the predictions and a uniform distribution.




\section{EXPERIMENTAL FRAMEWORK}

\subsection{Dataset}
{\color{black}For the primary task and disentanglement experiments for multimodal emotion recognition, we use the EmoVox dataset \cite{albanie2018emotion}}. The EmoVox dataset comprises of emotional labels on the VoxCeleb dataset obtained by predictions using a strong teacher network over eight emotional states: neutral, happiness, surprise, sadness, anger, disgust, fear and comtempt. Note that the teacher model was trained only using facial features (visual only). Overall, the dataset consists of interview videos from $1251$ celebrities spanning a wide range of ages and nationalities. For each video clip, we find the most dominant emotion based on the distribution and use that as our ground-truth label similar to \cite{albanie2018emotion}. 
The label distribution is heavily skewed towards a few emotion classes because emotions such as disgust, fear, contempt and surprise are rarely exhibited in interviews. Following previous approaches that deal with such imbalanced datasets \cite{busso2016msp}, we combine these labels into a single class `other`, resulting in $5$ emotion classes. Further, we discard videos corresponding to speakers belonging to the bottom $5$ percentile w.r.t the number of segments to reduce the imbalance in the number of speech segments per speaker.
We create three splits from the database: \textit{EmoVox-Train} to train models, \textit{EmoVox-Validation} for hyperparameter tuning, \textit{EmoVox-Test} to evaluate models on held out speech segments from speakers present in the train set. \textcolor{black}{The subset \textit{EmoVox-Train} corresponds to the \textit{Train} partition in \cite{albanie2018emotion}, whereas the \textit{EmoVox-Validation} and \textit{EmoVox-Test} were created from the \textit{Heard-Val} partition in \cite{albanie2018emotion}.}

\subsection{Experimental Settings}

The model architecture for the shared 2D Convolutional layers and the fully connected layers was adapted from \cite{nagrani2020disentangled} and modified to suit the dimensions of our inputs and outputs. \textcolor{black}{We use uniform duration videos of 12 seconds each as input to our system.}
\textcolor{black}{For the audio features we use $D_a=40$, and for the visual features we use $D_v=224$.} We fix the emotion embedding dimension to $2048$, while varying the speaker embedding dimension $2048$, $256$ and $64$. We use Adam optimizer with an initial learning rate of $1e-4$ and $1e-3$ for the primary branch and auxiliary branch updates respectively, decaying exponentially with a factor of $\gamma=0.9$.  For alternate training (Eqs. \ref{eq:eq1} and \ref{eq:eq2}), we chose $w_{em\_prim}$ and $w_{spk\_prim}$ to be $0.5$ each and $w_{em\_aux}$ and $w_{spk\_aux}$ to $0.3$ each. All parameters were chosen based on preliminary experiments on a subset of \textit{EmoVox-Train}. The emotion recognition performance was evaluated using unweighted F-score averaged across the $5$ emotion classes and for person identity with identification accuracy scores. Disentanglement is measured by combining both the F-score on emotion recognition using speaker embeddings and accuracy on person identification using emotion embeddings. Optimal models were chosen to give the best disentanglement (lowest score) on the \textit{EmoVox-Validation} set. All results are presented on the \textit{EmoVox-Test} set.

\section{RESULTS}
\label{sec:results}
\begin{figure*}[!htb]
	\centering
	\subfigure{\includegraphics[width=0.48\textwidth]{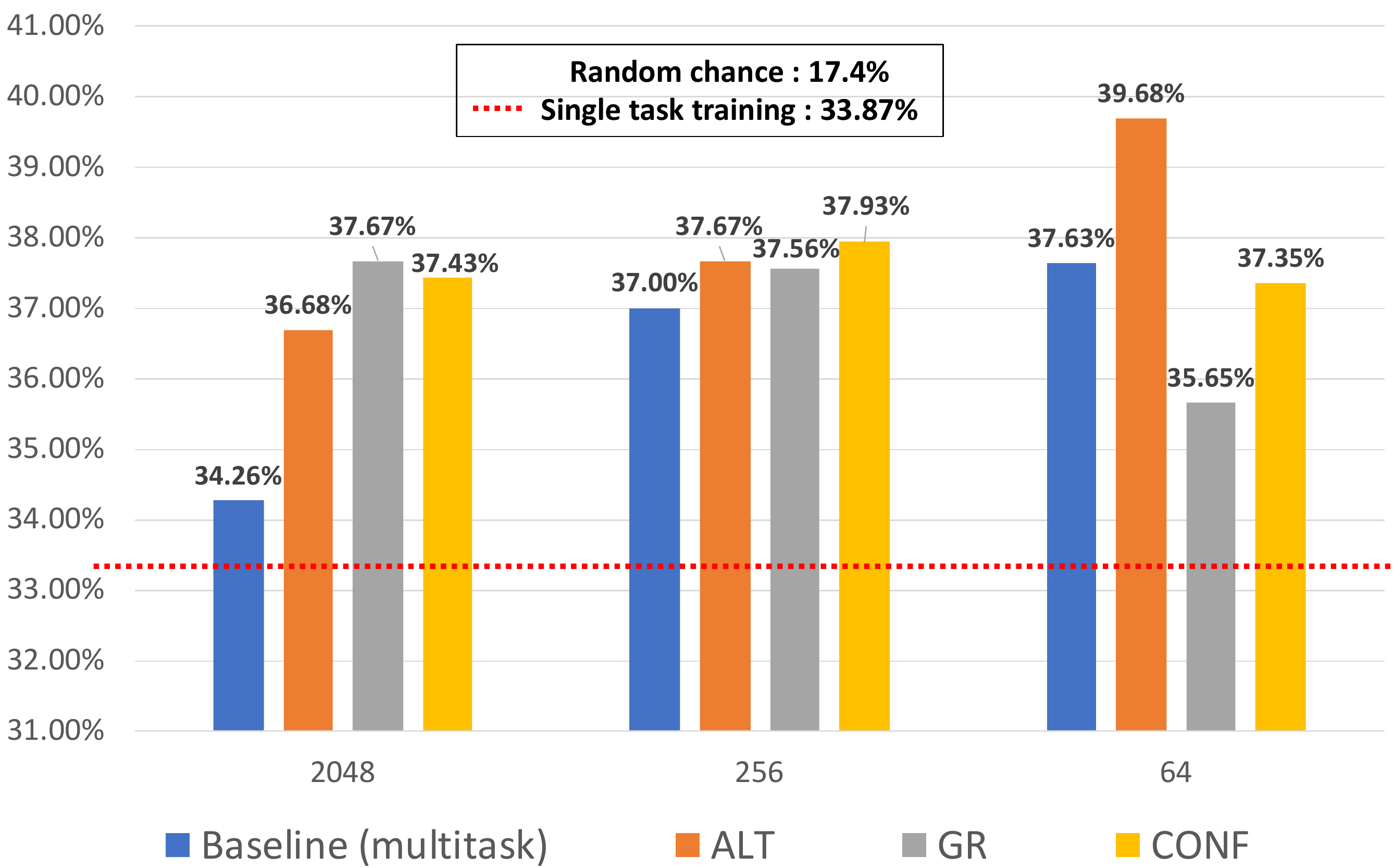} \label{fig:AER}}\hfill
	\subfigure{\includegraphics[width=0.48\textwidth]{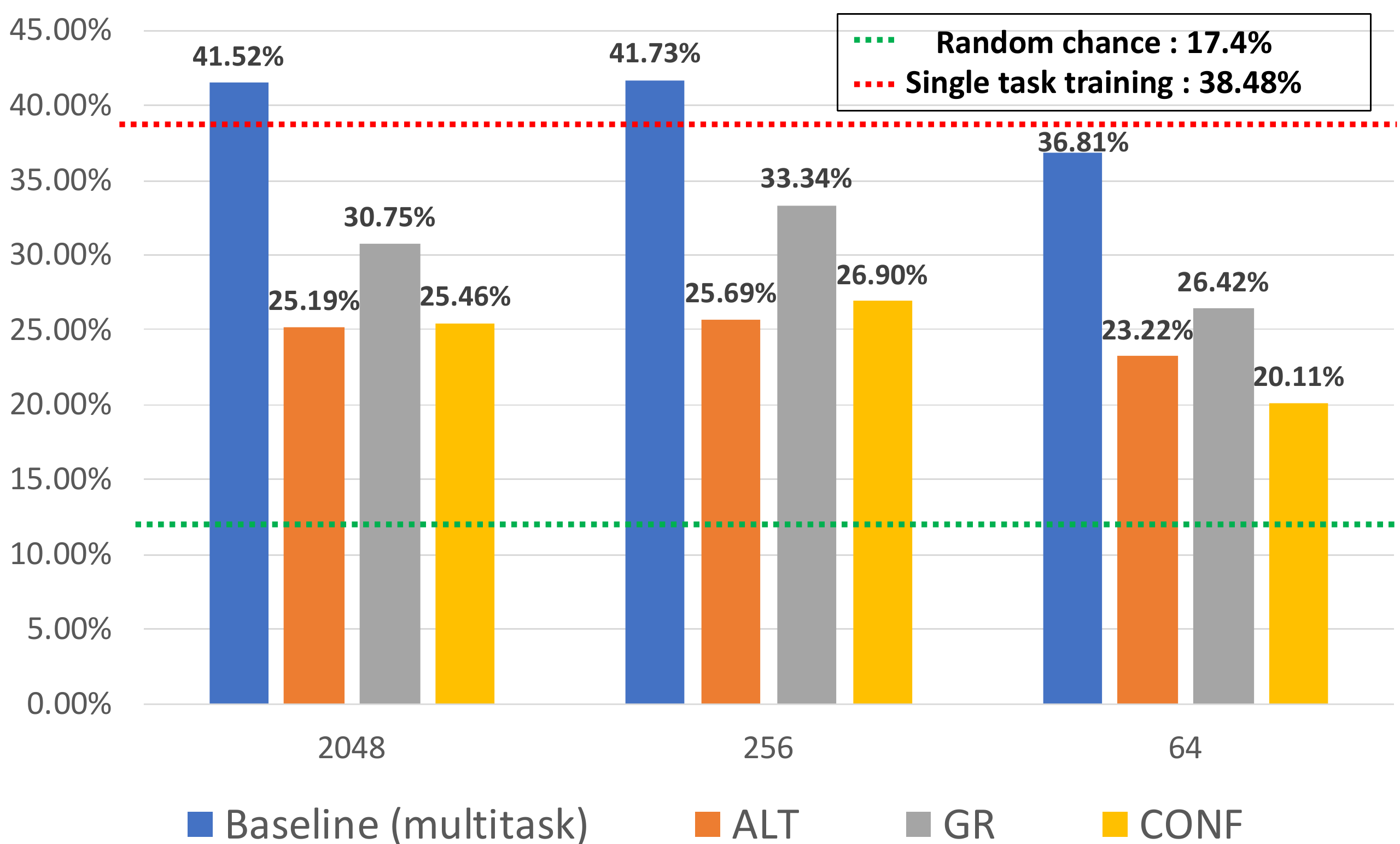} \label{fig:AER_Dis}}
	\caption{Unweighted Average F-scores for AER on \textit{EmoVox-Test} by varying speaker embedding dimension using (a) Emotion embeddings (Higher is better) (b) Speaker embeddings (Lower is better). A Stuart-Maxwell marginal homogeneity test comparing the results found, with statistical signficance, that all the models with disentanglement were different compared to the baseline model}
	
\end{figure*}

\subsection{Baseline models without disentanglement}
\textbf{Emotion Recognition:}
Figure~\ref{fig:AER} illustrates the primary emotion recognition results. The blue bars show the performance of all models trained using MTL and the dashed line shows the performance of Single-task learning~(STL) setup where the models are not trained on person identification. It is evident that MTL gives substantial gains in performance compared to STL setup. It is also observed that emotion recognition performance improves as the person identification embedding dimension is reduced, which may indicate better regularization with fewer embedding dimensions.\\
\textbf{Person identification:}
Table \ref{tab:SI} shows the person identification accuracy, trained with varying speaker embedding dimensions. It is worth noting that, despite the reduction in speaker embedding dimension, the models retain performance, pointing to the fact that the task of learning identity representations when both audio and visual modalities are available does not require many degrees of freedom.\\
\textbf{Identity information in emotion embeddings:}
Our preliminary experiments showed that the amount of person identity information entangled in emotion embeddings was minimal. Evaluating the person identification task using emotion embeddings produced an accuracy of 0.1\%, which was close to random chance performance. Therefore we focus on disentangling emotion information in identity embeddings.\\
\textbf{Emotion information in identity embeddings}
To baseline the amount of emotion information entangled in the speaker embeddings, we separately train single hidden layer neural network classifiers that predict the emotion labels from speaker embeddings. Figure \ref{fig:AER_Dis} illustrates the performance. First, it is worth noting that speaker embeddings from models trained for the single task of person identification retain substantial amount of emotion information, as shown by the red dashed line, compared to a random chance F-score of $17.40\%$ if all samples were predicted as `neutral` class ~(shown by the green dashed line). Further the blue bars illustrate the performance in the MTL setup where the F-scores are well above random chance as there is more information entanglement. This motivates the need for disentanglement to minimize the emotion information present in speaker embeddings without compromising performance on the emotion recognition, person identification tasks.

\begin{table}[t!]
	\centering
	\caption{Person identification accuracy (\%) comparing models with varying speaker embedding dimensions without and with disentanglement on \textit{EmoVox-Test}}
	\vspace{0.05in}
	\resizebox{0.44\textwidth}{!}{%
		\begin{tabular}{c|cccc}
			\toprule
			\begin{tabular}[c]{@{}c@{}}
				Emb Dim\end{tabular} & Baseline & ALT & GR & CONF \\ \midrule 
			2048 & 90.98 & 92.40 & 93.19 & 93.12 \\
			256 & 94.75 & 95.04 & 95.86 & 95.42 \\
			64 & 90.62 & 92.83 & 91.17 & 90.75 \\ 
			\bottomrule 
		\end{tabular}%
	\label{tab:SI}
	}
\end{table}

\subsection{Proposed models with disentanglement}
\label{ssec:disentangle}
Next we report the results of the proposed disentanglement techniques and compare them to the baseline models. We trained each disentanglement technique for all three configurations of speaker embedding dimension, $2048$, $256$ and $64$ to investigate their effect on disentanglement performance\\
\textbf{Emotion Recognition}
\label{sssec:AER_dis}
From Fig.~\ref{fig:AER}, we observe that models trained with all three disentanglement strategies outperform the baseline models trained without disentanglement in all but one case. In particular, ALT and CONF methods provide gains consistently across the various embedding dimensions. We performed a Stuart-Maxwell marginal homogeneity test comparing the results and found, with statistical signficance, that all the models with disentanglement were different compared to the baseline models \footnote{$H_0$: The predictions from the compared models are the same. Reject $H_0$ if $p<\alpha$ with $\alpha=0.01$ }. We also observe that, similar to the baseline models, models trained with disentanglement tend to perform better for reduced speaker embedding dimensions, though with smaller gains. \\
\textbf{Person identification}
Table \ref{tab:SI} shows the person identification accuracy for the models with disentanglement compared to the baseline without disentanglement. We observe that, in general, all models perform better after disentanglement when compared to the baseline without disentanglement. There is no clear evidence of one technique performing better than the other, though GR and ALT seem to perform marginally better compated to CONF. \\
\textbf{Emotion information in identity embeddings}
Fig.~\ref{fig:AER_Dis} illustrates the amount of emotion information in the person identity embeddings after explicit disentanglement. The drop in unweighted average F-score for emotion recognition shows the measure of obtained disentanglement. Compared to the models trained without disentanglement, we observe that the models trained with explicit disentanglement show reduction in F-score of predicting emotions from speaker embeddings. This is noticeable in all the three disentanglement techniques. ALT, CONF training show better disentanglement than GR. Overall, these results show the efficacy of using a separate auxiliary branch to disentangle the emotion information from speaker embeddings.
Furthermore, it can be observed that the models trained using the smallest speaker embedding dimension of $64$ shows the least amount of emotion information. This is expected because a reduced person identity embedding dimension creates a bottleneck to capture the primary identity information, and thus retains lesser amount of entangled emotion information. Considering the person identity dimension of 64, we see absolute gains of 2\% for emotion recognition while ALT training gives 13.5\% disentanglement.

\vspace{-1mm}
\section{CONCLUSIONS}
\vspace{-1mm}
This study analyses disentanglement techniques for emotion recognition in a multitask learning setup, where person identification is the secondary task. We showed with an audio-visual architecture that person identification helps emotion recognition performance. This comes at a cost, as there is significant information transfer between the tasks, which lets us predict emotional categories from speaker embeddings well above chance percentage. To combat this we studied three disentanglement techniques, each reducing the amount of information that is entangled while maintaining or improving performance on the primary task. For our next steps we will explore and validate these methods on other databases which have stronger emotion labels. Furthermore, it is of interest to dig deeper into the reasons for differences in performance across the various disentanglement methods. Finally, this paper shows that there is significant emotional information in the speaker embeddings and the contrary is not necessarily true. Therefore we will explore a hierarchical structure where emotion recognition is more downstream than the person identification task.

\renewcommand{\normalsize}{\fontsize{9}{10.5}\selectfont}
\normalsize
\bibliographystyle{IEEEbib}
\bibliography{refs}

\begin{thebibliography}{10}

\bibitem{pantic2003toward}
M.~Pantic and J.~M.~L. Rothkrantz,
\newblock ``Toward an affect-sensitive multimodal human-computer interaction,''
\newblock {\em Proceedings of the IEEE}, vol. 91, no. 9, pp. 1370--1390, 2003.

\bibitem{mehrabian2008communication}
A.~Mehrabian,
\newblock ``Communication without words,''
\newblock {\em Communication theory}, vol. 6, pp. 193--200, 2008.

\bibitem{kim2013deep}
Y.~Kim, H.~Lee, and E.~M. Provost,
\newblock ``Deep learning for robust feature generation in audiovisual emotion
  recognition,''
\newblock in {\em 2013 IEEE international conference on acoustics, speech and
  signal processing}, Vancouver, BC, Canada, 2013, IEEE, pp. 3687--3691.

\bibitem{wang2008recognizing}
Y.~Wang and L.~Guan,
\newblock ``Recognizing human emotional state from audiovisual signals,''
\newblock {\em IEEE transactions on multimedia}, vol. 10, no. 5, pp. 936--946,
  2008.

\bibitem{song2004audio}
M.~Song, J.~Bu, C.~Chen, and N.~Li,
\newblock ``Audio-visual based emotion recognition-a new approach,''
\newblock in {\em Proceedings of the 2004 IEEE Computer Society Conference on
  Computer Vision and Pattern Recognition, 2004. CVPR 2004.}, Washington, DC,
  USA, 2004, IEEE, vol.~2, pp. II--II.

\bibitem{parthasarathy2017jointly}
S.~Parthasarathy and C.~Busso,
\newblock ``Jointly predicting arousal, valence and dominance with multi-task
  learning.,''
\newblock in {\em Interspeech}, Stockholm, Sweden, 2017, pp. 1103--1107.

\bibitem{li2019improved}
Y.~Li, T.~Zhao, and T.~Kawahara,
\newblock ``Improved end-to-end speech emotion recognition using self attention
  mechanism and multitask learning.,''
\newblock in {\em Interspeech}, Graz, Austria, 2019, pp. 2803--2807.

\bibitem{kim2017towards}
J.~Kim, G.~Englebienne, P.~K. Truong, and V.~Evers,
\newblock ``Towards speech emotion recognition" in the wild" using aggregated
  corpora and deep multi-task learning,''
\newblock {\em arXiv preprint arXiv:1708.03920}, 2017.

\bibitem{zhang2017cross}
B.~Zhang, E.~M. Provost, and G.~Essl,
\newblock ``Cross-corpus acoustic emotion recognition with multi-task learning:
  Seeking common ground while preserving differences,''
\newblock {\em IEEE Transactions on Affective Computing}, vol. 10, no. 1, pp.
  85--99, 2017.

\bibitem{liang2020model}
J.~Liang, Z.~Liu, J.~Zhou, X.~Jiang, C~.Zhang, and F.~Wang,
\newblock ``Model-protected multi-task learning,''
\newblock {\em IEEE Transactions on Pattern Analysis and Machine Intelligence},
  pp. 1--1, 2020.

\bibitem{xiao2018learning}
L.~Xiao, H.~Zhang, W.~Chen, Y.~Wang, and Y.~Jin,
\newblock ``Learning what to share: Leaky multi-task network for text
  classification,''
\newblock in {\em Proceedings of the 27th International Conference on
  Computational Linguistics}, Santa Fe, NM, USA, 2018, pp. 2055--2065.

\bibitem{williams2019disentangling}
J.~Williams and S.~King,
\newblock ``Disentangling style factors from speaker representations.,''
\newblock in {\em INTERSPEECH}, Graz, Austria, 2019, pp. 3945--3949.

\bibitem{chaplin2015gender}
T.~M. Chaplin,
\newblock ``Gender and emotion expression: A developmental contextual
  perspective,''
\newblock {\em Emotion Review}, vol. 7, no. 1, pp. 14--21, 2015.

\bibitem{alvi2018turning}
M.~Alvi, A.~Zisserman, and C.~Nell{\aa}ker,
\newblock ``Turning a blind eye: Explicit removal of biases and variation from
  deep neural network embeddings,''
\newblock in {\em Proceedings of the European Conference on Computer Vision
  (ECCV)}, Munich, Germany, 2018, pp. 556--572.

\bibitem{abdelwahab2018domain}
M.~Abdelwahab and C.~Busso,
\newblock ``Domain adversarial for acoustic emotion recognition,''
\newblock {\em IEEE/ACM Transactions on Audio, Speech, and Language
  Processing}, vol. 26, no. 12, pp. 2423--2435, 2018.

\bibitem{mao2017learning}
Q.~Mao, G.~Xu, W.~Xue, J.~Gou, and Y.~Zhan,
\newblock ``Learning emotion-discriminative and domain-invariant features for
  domain adaptation in speech emotion recognition,''
\newblock {\em Speech Communication}, vol. 93, pp. 1--10, 2017.

\bibitem{tu2019towards}
M.~Tu, Y.~Tang, J.~Huang, X.~He, and B.~Zhou,
\newblock ``Towards adversarial learning of speaker-invariant representation
  for speech emotion recognition,''
\newblock {\em arXiv preprint arXiv:1903.09606}, 2019.

\bibitem{li2020speaker}
H.~Li, M.~Tu, J.~Huang, S.~Narayanan, and P.~Georgiou,
\newblock ``Speaker-invariant affective representation learning via adversarial
  training,''
\newblock in {\em ICASSP 2020-2020 IEEE International Conference on Acoustics,
  Speech and Signal Processing (ICASSP)}, Barcelona, Spain, 2020, IEEE, pp.
  7144--7148.

\bibitem{kang2020disentangled}
W.~H. Kang, S.~H. Mun, M.~H. Han, and N.~S. Kim,
\newblock ``Disentangled speaker and nuisance attribute embedding for robust
  speaker verification,''
\newblock {\em IEEE Access}, vol. 8, pp. 141838--141849, 2020.

\bibitem{jaiswal2020privacy}
M.~Jaiswal and E.~M. Provost,
\newblock ``Privacy enhanced multimodal neural representations for emotion
  recognition.,''
\newblock in {\em AAAI}, 2020, pp. 7985--7993.

\bibitem{parthasarathy2017study}
S.~Parthasarathy, C.~Zhang, J.~H.~L. Hansen, and C.~Busso,
\newblock ``A study of speaker verification performance with expressive
  speech,''
\newblock in {\em 2017 IEEE International Conference on Acoustics, Speech and
  Signal Processing (ICASSP)}, New Orleans, LA, USA, 2017, IEEE, pp.
  5540--5544.

\bibitem{wu2006study}
W.~Wu, T.~F. Zheng, M.~X. Xu, and H.~J. Bao,
\newblock ``Study on speaker verification on emotional speech,''
\newblock in {\em Ninth International Conference on Spoken Language
  Processing}, Pittsburgh, PA, USA, 2006, pp. 2102--2105.

\bibitem{bao2007emotion}
H.~Bao, M.~X. Xu, and T.~F. Zheng,
\newblock ``Emotion attribute projection for speaker recognition on emotional
  speech,''
\newblock in {\em Eighth Annual Conference of the International Speech
  Communication Association}, Antwerp, Belgium, 2007, pp. 758--761.

\bibitem{krothapalli2012neural}
S.~R. Krothapalli, J.~Yadav, S.~Sarkar, G.~S. Koolagudi, and A.~K. Vuppala,
\newblock ``Neural network based feature transformation for emotion independent
  speaker identification,''
\newblock {\em International Journal of Speech Technology}, vol. 15, no. 3, pp.
  335--349, 2012.

\bibitem{nagrani2020disentangled}
A.~Nagrani, J.~S~. Chung, S.~Albanie, and A.~Zisserman,
\newblock ``Disentangled speech embeddings using cross-modal
  self-supervision,''
\newblock in {\em ICASSP 2020-2020 IEEE International Conference on Acoustics,
  Speech and Signal Processing (ICASSP)}, Barcelona, Spain, 2020, IEEE, pp.
  6829--6833.

\bibitem{ganin2015unsupervised}
Y.~Ganin and V.~Lempitsky,
\newblock ``Unsupervised domain adaptation by backpropagation,''
\newblock in {\em International conference on machine learning}, Lille, France,
  2015, PMLR, pp. 1180--1189.

\bibitem{goodfellow2014generative}
I.~Goodfellow, J.~Pouget-Abadie, M.~Mirza, B.~Xu, D.~Warde-Farley, S.~Ozair,
  A.~Courville, and Y.~Bengio,
\newblock ``Generative adversarial nets,''
\newblock in {\em Advances in neural information processing systems}, Montreal,
  Canada, 2014, pp. 2672--2680.

\bibitem{tzeng2015simultaneous}
E.~Tzeng, J.~Hoffman, T.~Darrell, and K.~Saenko,
\newblock ``Simultaneous deep transfer across domains and tasks,''
\newblock in {\em Proceedings of the IEEE International Conference on Computer
  Vision}, Santiago, Chile, pp. 4068--4076.

\bibitem{albanie2018emotion}
S.~Albanie, A.~Nagrani, A.~Vedaldi, and A.~Zisserman,
\newblock ``Emotion recognition in speech using cross-modal transfer in the
  wild,''
\newblock in {\em Proceedings of the 26th ACM international conference on
  Multimedia}, Seoul, Republic of Korea, pp. 292--301.

\bibitem{busso2016msp}
C.~Busso, S.~Parthasarathy, A.~Burmania, M.~AbdelWahab, N.~Sadoughi, and E.~M.
  Provost,
\newblock ``Msp-improv: An acted corpus of dyadic interactions to study emotion
  perception,''
\newblock {\em IEEE Transactions on Affective Computing}, vol. 8, no. 1, pp.
  67--80, 2016.

\end{thebibliography}

\end{document}